\documentclass[prd,aps,showpacs,epsf,floats]{revtex4}
\usepackage{amsfonts}
\usepackage{amsmath}

\input{tcilatex}

\begin{document}

\title{Information Geometry and Chaos on Negatively Curved Statistical
Manifolds}
\author{Carlo Cafaro${}$}
\email{carlocafaro2000@yahoo.it}
\affiliation{Department of Physics, State University of New York at Albany-SUNY,1400
Washington\\
Avenue, Albany, NY 12222, USA}

\begin{abstract}
A novel information-geometric approach to chaotic dynamics on curved
statistical manifolds based on Entropic Dynamics (ED) is suggested.
Furthermore, an information-geometric analogue of the Zurek-Paz quantum
chaos criterion is proposed. It is shown that the hyperbolicity of a
non-maximally symmetric $6N$-dimensional statistical manifold $\mathcal{M}%
_{s}$ underlying an ED Gaussian model describing an arbitrary system of $3N$
non-interacting degrees of freedom leads to linear information-geometric
entropy growth and to exponential divergence of the Jacobi vector field
intensity, quantum and classical features of chaos respectively.
\end{abstract}

\maketitle



\section{\textbf{INTRODUCTION}}

Entropic Dynamics (ED) $\left[ 1\right] $, namely the combination of
principles of inductive inference (Maximum relative Entropy Methods, $\left[
2\right] $) and Information Geometry (IG) $\left[ 3\right] $, is a
theoretical framework constructed on statistical manifolds and it is
developed to investigate the possibility that laws of physics, either
classical or quantum, might reflect laws of inference rather than laws of
nature. This paper is a follow up of a series of the author's works $\left[ 4%
\text{, }5\right] $. In this article, we use the ED theoretical framework to
explore the possibility of constructing a unifying (classical and quantum)
criterion of chaos. We assume the system under investigation has $3N$
degrees of freedom, each one described by two pieces of relevant
information, its expectation value and its variance (Gaussian statistical
variables). This leads to consider an ED model on a non-maximally symmetric $%
6N$- dimensional statistical manifold $\mathcal{M}_{s}$. The manifold $%
\mathcal{M}_{s}$ has constant negative Ricci curvature proportional to the
number of degrees of freedom of the system, $R_{\mathcal{M}_{s}}=-3N$. An
information-geometric analog of the Zurek-Paz quantum chaos criterion is
suggested. It is shown that the system explores statistical volume elements
on $\mathcal{M}_{s}$ at an exponential rate. We define a dynamical
information-geometric entropy $S_{\mathcal{M}_{s}}$ \ of the system and we
show it increases linearly in time (statistical evolution parameter) and is
proportional to the number of degrees of freedom of the system. The
geodesics on $\mathcal{M}_{s}$ are hyperbolic trajectories. Using the
Jacobi-Levi-Civita (JLC) equation for geodesic spread, it is shown that the
Jacobi vector field intensity $J_{\mathcal{M}_{s}}$ diverges exponentially
and is proportional to the number of degrees of freedom of the system. Thus, 
$R_{\mathcal{M}_{s}}$, $S_{\mathcal{M}_{s}}$ and $J_{\mathcal{M}_{s}}$ are
proportional to the number of Gaussian-distributed microstates of the
system. This proportionality leads to conclude there exists a substantial
link among these information-geometric indicators of chaoticity.

\section{THE\ ED\ GAUSSIAN\ MODEL}

Given two probability distributions, how can one define a notion of
"distance" between them? The answer to this question is provided by IG. As
it is shown in $[6]$ and $\left[ 7\right] $, the notion of distance between
dissimilar probability distributions is quantified by the Fisher-Rao
information metric tensor. We consider an ED model whose microstates span a $%
3N$-dimensional space labelled by the variables $\left\{ \vec{X}\right\}
=\left\{ \vec{x}^{\left( 1\right) }\text{, }\vec{x}^{\left( 2\right) }\text{%
,...., }\vec{x}^{\left( N\right) }\right\} $ with $\vec{x}^{\left( \alpha
\right) }\equiv \left( x_{1}^{\left( \alpha \right) }\text{, }x_{2}^{\left(
\alpha \right) }\text{, }x_{3}^{\left( \alpha \right) }\right) $, $\alpha =1$%
,...., $N$ and $x_{a}^{\left( \alpha \right) }\in 
\mathbb{R}
$ with $a=1$, $2$, $3$ . We assume the only testable information pertaining
to the quantities $x_{a}^{\left( \alpha \right) }$ consists of the
expectation values $\left\langle x_{a}^{\left( \alpha \right) }\right\rangle 
$ and the variance $\Delta x_{a}^{\left( \alpha \right) }=\sqrt{\left\langle
\left( x_{a}^{\left( \alpha \right) }-\left\langle x_{a}^{\left( \alpha
\right) }\right\rangle \right) ^{2}\right\rangle }$. The set of these
expected values define the $6N$-dimensional space of macrostates of the
system. A measure of distinguishability among the states of the ED model is
achieved by assigning a probability distribution $P\left( \vec{X}\left\vert 
\vec{\Theta}\right. \right) $ to each macrostate $\vec{\Theta}$ where $%
\left\{ \vec{\Theta}\right\} =\left\{ ^{\left( 1\right) }\theta _{a}^{\left(
\alpha \right) }\text{,}^{\left( 2\right) }\theta _{a}^{\left( \alpha
\right) }\right\} $ with $\alpha =1$, $2$,$....$, $N$ and $a=1,2,3$. The
process of assigning a probability distribution to each state provides $%
\mathcal{M}_{S}$ with a metric structure. Specifically, the Fisher-Rao
information metric defined in $(7)$ is a measure of distinguishability among
macrostates. It assigns an IG to the space of states.

\section{THE\ STATISTICAL\ MANIFOLD\ $\mathcal{M}_{S}$}

Consider an arbitrary physical system evolving over a $3N$-dimensional
space.\ The variables $\left\{ \vec{X}\right\} =\left\{ \vec{x}^{\left(
1\right) }\text{, }\vec{x}^{\left( 2\right) }\text{,...., }\vec{x}^{\left(
N\right) }\right\} $ label the $3N$-dimensional space of microstates of the
system. Each macrostate may be thought as a point of a $6N$-dimensional
statistical manifold with coordinates given by the numerical values of the
expectations $^{\left( 1\right) }\theta _{a}^{\left( \alpha \right)
}=\left\langle x_{a}^{\left( \alpha \right) }\right\rangle $ and $^{\left(
2\right) }\theta _{a}^{\left( \alpha \right) }=\Delta x_{a}^{\left( \alpha
\right) }\equiv \sqrt{\left\langle \left( x_{a}^{\left( \alpha \right)
}-\left\langle x_{a}^{\left( \alpha \right) }\right\rangle \right)
^{2}\right\rangle }$. The available information can be written in the form
of the following $6N$ information constraint equations,%
\begin{equation}
\begin{array}{c}
\left\langle x_{a}^{\left( \alpha \right) }\right\rangle
=\dint\limits_{-\infty }^{+\infty }dx_{a}^{\left( \alpha \right)
}x_{a}^{\left( \alpha \right) }P_{a}^{\left( \alpha \right) }\left(
x_{a}^{\left( \alpha \right) }\left\vert ^{\left( 1\right) }\theta
_{a}^{\left( \alpha \right) }\text{,}^{\left( 2\right) }\theta _{a}^{\left(
\alpha \right) }\right. \right) \\ 
\\ 
\Delta x_{a}^{\left( \alpha \right) }=\left[ \dint\limits_{-\infty
}^{+\infty }dx_{a}^{\left( \alpha \right) }\left( x_{a}^{\left( \alpha
\right) }-\left\langle x_{a}^{\left( \alpha \right) }\right\rangle \right)
^{2}P_{a}^{\left( \alpha \right) }\left( x_{a}^{\left( \alpha \right)
}\left\vert ^{\left( 1\right) }\theta _{a}^{\left( \alpha \right) }\text{,}%
^{\left( 2\right) }\theta _{a}^{\left( \alpha \right) }\right. \right) %
\right] ^{\frac{1}{2}}%
\end{array}%
\end{equation}%
where $^{\left( 1\right) }\theta _{a}^{\left( \alpha \right) }=\left\langle
x_{a}^{\left( \alpha \right) }\right\rangle $ and $^{\left( 2\right) }\theta
_{a}^{\left( \alpha \right) }=\Delta x_{a}^{\left( \alpha \right) }$ with $%
\alpha =1$, $2$,$....$, $N$ and $a=1,2,3$. The probability distributions $%
P_{a}^{\left( \alpha \right) }$ are constrained by the conditions of
normalization,%
\begin{equation}
\dint\limits_{-\infty }^{+\infty }dx_{a}^{\left( \alpha \right)
}P_{a}^{\left( \alpha \right) }\left( x_{a}^{\left( \alpha \right)
}\left\vert ^{\left( 1\right) }\theta _{a}^{\left( \alpha \right) }\text{,}%
^{\left( 2\right) }\theta _{a}^{\left( \alpha \right) }\right. \right) =1%
\text{.}
\end{equation}%
Information theory identifies the Gaussian distribution as the maximum
entropy distribution if only the expectation value and the variance are
known. The distribution that best reflects the information contained in the
prior distribution $m\left( \vec{X}\right) $ updated by the information $%
\left( \left\langle x_{a}^{\left( \alpha \right) }\right\rangle ,\Delta
x_{a}^{\left( \alpha \right) }\right) $ is obtained by maximizing the
relative entropy 
\begin{equation}
S\left( \vec{\Theta}\right) =-\int d^{3N}\vec{X}P\left( \vec{X}\left\vert 
\vec{\Theta}\right. \right) \log \left( \frac{P\left( \vec{X}\left\vert \vec{%
\Theta}\right. \right) }{m\left( \vec{X}\right) }\right) \text{,}
\end{equation}%
where $m(\vec{X})$ is the prior probability distribution. As a working
hypothesis, the prior $m\left( \vec{X}\right) $ is set to be uniform since
we assume the lack of prior available information about the system
(postulate of equal \textit{a priori} probabilities). Upon maximizing $%
\left( 3\right) $, given the constraints $\left( 1\right) $ and $\left(
2\right) $, we obtain%
\begin{equation}
P\left( \vec{X}\left\vert \vec{\Theta}\right. \right) =\dprod\limits_{\alpha
=1}^{N}\dprod\limits_{a=1}^{3}P_{a}^{\left( \alpha \right) }\left(
x_{a}^{\left( \alpha \right) }\left\vert \mu _{a}^{\left( \alpha \right) }%
\text{, }\sigma _{a}^{\left( \alpha \right) }\right. \right)
\end{equation}%
where%
\begin{equation}
P_{a}^{\left( \alpha \right) }\left( x_{a}^{\left( \alpha \right)
}\left\vert \mu _{a}^{\left( \alpha \right) }\text{, }\sigma _{a}^{\left(
\alpha \right) }\right. \right) =\left( 2\pi \left[ \sigma _{a}^{\left(
\alpha \right) }\right] ^{2}\right) ^{-\frac{1}{2}}\exp \left[ -\frac{\left(
x_{a}^{\left( \alpha \right) }-\mu _{a}^{\left( \alpha \right) }\right) ^{2}%
}{2\left( \sigma _{a}^{\left( \alpha \right) }\right) ^{2}}\right]
\end{equation}%
and $^{\left( 1\right) }\theta _{a}^{\left( \alpha \right) }=\mu
_{a}^{\left( \alpha \right) }$, $^{\left( 2\right) }\theta _{a}^{\left(
\alpha \right) }=\sigma _{a}^{\left( \alpha \right) }$. The probability
distribution $(4)$ encodes the available information concerning the system.
Note that we have assumed uncoupled constraints among microvariables $%
x_{a}^{\left( \alpha \right) }$. In other words, we assumed that information
about correlations between the microvariables need not to be tracked. This
assumption leads to the simplified product rule $\left( 4\right) $. However,
coupled constraints would lead to a generalized product rule in $\left(
4\right) $ and to a metric tensor $\left( 7\right) $ with non-trivial
off-diagonal elements (covariance terms). Such generalizations would require
more delicate analysis. Deviations from Gaussian-type information
constraints and the presence of a nonvanishing correlation coefficient $%
\left( r=\frac{\left\langle \left( x_{1}-\mu _{1}\right) \left( x_{2}-\mu
_{2}\right) \right\rangle }{\sigma _{1}\sigma _{2}}\text{, }r\text{ is the
correlation coefficient of the two dependent random variables }x_{1}\text{%
and }x_{2}\right) $ among the random microvariables $x_{a}^{\left( \alpha
\right) }$ are some of the new topics appearing in a forthcoming paper $%
\left[ 8\right] $.

\subsection{METRIC\ STRUCTURE\ OF\ $\mathcal{M}_{S}$}

The dimensionless line element $ds$ between $P\left( \vec{X}\left\vert \vec{%
\Theta}\right. \right) $ and $P\left( \vec{X}\left\vert \vec{\Theta}+d\vec{%
\Theta}\right. \right) $ is given by,%
\begin{equation}
ds^{2}=g_{\mu \nu }d\Theta ^{\mu }d\Theta ^{\nu }
\end{equation}%
where%
\begin{equation}
g_{\mu \nu }=\int d\vec{X}P\left( \vec{X}\left\vert \vec{\Theta}\right.
\right) \frac{\partial \log P\left( \vec{X}\left\vert \vec{\Theta}\right.
\right) }{\partial \Theta ^{\mu }}\frac{\partial \log P\left( \vec{X}%
\left\vert \vec{\Theta}\right. \right) }{\partial \Theta ^{\nu }}
\end{equation}%
is the Fisher-Rao metric. Substituting $\left( 4\right) $ into $\left(
7\right) $, the metric $g_{\mu \nu }$ on $\mathcal{M}_{s}$ becomes a $%
6N\times 6N$ matrix $M$ made up of $3N$ blocks $M_{2\times 2}$ with
dimension $2\times 2$ given by,%
\begin{equation}
M_{2\times 2}=\left( 
\begin{array}{cc}
\left( \sigma _{a}^{\left( \alpha \right) }\right) ^{-2} & 0 \\ 
0 & 2\times \left( \sigma _{a}^{\left( \alpha \right) }\right) ^{-2}%
\end{array}%
\right)
\end{equation}%
with $\alpha =1$, $2$,$....$, $N$ and $a=1,2,3$. From $\left( 7\right) $,
the "length" element $\left( 6\right) $ reads,%
\begin{equation}
ds^{2}=\dsum\limits_{\alpha =1}^{N}\dsum\limits_{a=1}^{3}\left[ \frac{1}{%
\left( \sigma _{a}^{\left( \alpha \right) }\right) ^{2}}d\mu _{a}^{\left(
\alpha \right) 2}+\frac{2}{\left( \sigma _{a}^{\left( \alpha \right)
}\right) ^{2}}d\sigma _{a}^{\left( \alpha \right) 2}\right] \text{.}
\end{equation}%
We bring attention to the fact that the metric structure of $\mathcal{M}_{s}$
is an emergent (not fundamental) structure. It arises only after assigning a
probability distribution $P\left( \vec{X}\left\vert \vec{\Theta}\right.
\right) $ to each state $\vec{\Theta}$.

\subsection{CURVATURE\ OF\ $\mathcal{M}_{s}$}

The Ricci scalar curvature $R$ is given by,%
\begin{equation}
R=g^{\mu \nu }R_{\mu \nu }\text{,}
\end{equation}%
where $g^{\mu \nu }g_{\nu \rho }=\delta _{\rho }^{\mu }$ so that $g^{\mu \nu
}=\left( g_{\mu \nu }\right) ^{-1}$. The Ricci tensor $R_{\mu \nu }$ is
given by,%
\begin{equation}
R_{\mu \nu }=\partial _{\varepsilon }\Gamma _{\mu \nu }^{\varepsilon
}-\partial _{\nu }\Gamma _{\mu \varepsilon }^{\varepsilon }+\Gamma _{\mu \nu
}^{\varepsilon }\Gamma _{\varepsilon \eta }^{\eta }-\Gamma _{\mu \varepsilon
}^{\eta }\Gamma _{\nu \eta }^{\varepsilon }\text{.}
\end{equation}%
The Christoffel symbols $\Gamma _{\mu \nu }^{\rho }$ appearing in the Ricci
tensor are defined in the standard way, 
\begin{equation}
\Gamma _{\mu \nu }^{\rho }=\frac{1}{2}g^{\rho \varepsilon }\left( \partial
_{\mu }g_{\varepsilon \nu }+\partial _{\nu }g_{\mu \varepsilon }-\partial
_{\varepsilon }g_{\mu \nu }\right) .
\end{equation}%
Using $\left( 9\right) $ and the definitions given above, we can show that
the Ricci scalar curvature becomes%
\begin{equation}
R_{\mathcal{M}_{s}}=-3N<0\text{.}
\end{equation}%
From $(13)$ we conclude that $\mathcal{M}_{s}$ is a $6N$-dimensional
statistical manifold of constant negative Ricci scalar curvature. A detailed
analysis on the calculation of Christoffel connection coefficients using the
ED formalism for a four-dimensional manifold of Gaussians can be found in $%
\left[ 5\right] $. Furthermore, it can be shown that $\mathcal{M}_{s}$ is
not a pseudosphere (maximally symmetric manifold) since its sectional
curvature is not constant. Considerations about the negativity of the Ricci
curvature as a \textit{strong criterion} of dynamical instability and the
necessity of \textit{compactness} of $\mathcal{M}_{s}$\textit{\ }in "true"
chaotic dynamical systems will appear in $\left[ 8\right] $.

\section{CANONICAL\ FORMALISM FOR THE ED-GAUSSIAN MODEL}

At this point, we study the trajectories of the system on $\mathcal{M}_{s}$.
We emphasize ED can be derived from a standard principle of least action
(Maupertuis- Euler-Lagrange-Jacobi-type) $\left[ 1\text{, }9\right] $. The
geodesic equations for the macrovariables of the Gaussian ED model are given
by,%
\begin{equation}
\frac{d^{2}\Theta ^{\mu }}{d\tau ^{2}}+\Gamma _{\nu \rho }^{\mu }\frac{%
d\Theta ^{\nu }}{d\tau }\frac{d\Theta ^{\rho }}{d\tau }=0
\end{equation}%
with $\mu =1$, $2$,...,$6N$. Observe that the geodesic equations are\textit{%
\ nonlinear}, second order coupled ordinary differential equations.

\subsection{GEODESICS ON $\mathcal{M}_{s}$}

We seek the explicit form of $(14)$ for the pairs of statistical coordinates 
$(\mu _{a}^{\left( \alpha \right) }$, $\sigma _{a}^{\left( \alpha \right) })$%
. Substituting the explicit expression of the Christoffel connection
coefficients into $\left( 14\right) $, the geodesic equations for the
macrovariables $\mu _{a}^{\left( \alpha \right) }$ and $\sigma _{a}^{\left(
\alpha \right) }$ associated to the microstate $x_{a}^{\left( \alpha \right)
}$ become,%
\begin{equation}
\text{ }\frac{d^{2}\mu _{a}^{\left( \alpha \right) }}{d\tau ^{2}}-\frac{2}{%
\sigma _{a}^{\left( \alpha \right) }}\frac{d\mu _{a}^{\left( \alpha \right) }%
}{d\tau }\frac{d\sigma _{a}^{\left( \alpha \right) }}{d\tau }=0\text{, }%
\frac{d^{2}\sigma _{a}^{\left( \alpha \right) }}{d\tau ^{2}}-\frac{1}{\sigma
_{a}^{\left( \alpha \right) }}\left( \frac{d\sigma _{a}^{\left( \alpha
\right) }}{d\tau }\right) ^{2}+\frac{1}{2\sigma _{a}^{\left( \alpha \right) }%
}\left( \frac{d\mu _{a}^{\left( \alpha \right) }}{d\tau }\right) ^{2}=0\text{%
.}
\end{equation}%
with $\alpha =1$, $2$,$....$, $N$ and $a=1,2,3$. This is a set of coupled
ordinary differential equations, whose solutions are%
\begin{equation}
\begin{array}{c}
\mu _{a}^{\left( \alpha \right) }\left( \tau \right) =\frac{\left(
B_{a}^{\left( \alpha \right) }\right) ^{2}}{2\beta _{a}^{\left( \alpha
\right) }}\frac{1}{\cosh \left( 2\beta _{a}^{\left( \alpha \right) }\tau
\right) -\sinh \left( 2\beta _{a}^{\left( \alpha \right) }\tau \right) +%
\frac{\left( B_{a}^{\left( \alpha \right) }\right) ^{2}}{8\left( \beta
_{a}^{\left( \alpha \right) }\right) ^{2}}}+C_{a}^{\left( \alpha \right) }%
\text{,} \\ 
\\ 
\sigma _{a}^{\left( \alpha \right) }\left( \tau \right) =B_{a}^{\left(
\alpha \right) }\frac{\cosh \left( \beta _{a}^{\left( \alpha \right) }\tau
\right) -\sinh \left( \beta _{a}^{\left( \alpha \right) }\tau \right) }{%
\cosh \left( 2\beta _{a}^{\left( \alpha \right) }\tau \right) -\sinh \left(
2\beta _{a}^{\left( \alpha \right) }\tau \right) +\frac{\left( B_{a}^{\left(
\alpha \right) }\right) ^{2}}{8\left( \beta _{a}^{\left( \alpha \right)
}\right) ^{2}}}\text{.}%
\end{array}%
\end{equation}%
The quantities $B_{a}^{\left( \alpha \right) }$, $C_{a}^{\left( \alpha
\right) }$, $\beta _{a}^{\left( \alpha \right) }$ are \textit{real}
integration constants and they can be evaluated once the boundary conditions
are specified. We are interested in investigating the stability of the
trajectories of the ED model considered on $\mathcal{M}_{s}$. It is known $%
\left[ 9\right] $ that the Riemannian curvature of a manifold is closely
connected with the behavior of the geodesics on it. If the Riemannian
curvature of a manifold is negative, geodesics (initially parallel) rapidly
diverge from one another. For the sake of simplicity, we assume very special
initial conditions: $B_{a}^{\left( \alpha \right) }\equiv \Lambda $, $\beta
_{a}^{\left( \alpha \right) }\equiv \lambda \in 
\mathbb{R}
^{+}$, $C_{a}^{\left( \alpha \right) }=0$, $\forall \alpha =1$, $2$,$....$, $%
N$ and $a=1,2,3$ . However, the conclusion we reach can be generalized to
more arbitrary initial conditions. It is worthwhile noticing that, in our
case, $\mathcal{M}_{s}$ is a geodesically complete manifold since every
maximal geodesic is well-defined for all temporal parameters $\tau $.
Therefore, $\mathcal{M}_{s}$ represents a natural setting for \textit{global}
questions in this Riemannian geometric framework applied to probability
theory and the search for a \textit{weak criterion} of chaos can be carried
out $\left[ 8\right] $.

\section{LINEARITY OF THE INFORMATION-GEOMETRIC DYNAMICAL ENTROPY}

Recall that $\mathcal{M}_{s}$ is the space of probability distributions $%
P\left( \vec{X}\left\vert \vec{\Theta}\right. \right) $ labeled by $6N$
statistical parameters $\vec{\Theta}$. These parameters are the coordinates
for the point $P$, and in these coordinates a volume element $dV_{\mathcal{M}%
_{s}}$ reads, 
\begin{equation}
dV_{\mathcal{M}_{S}}=\sqrt{g}d^{6N}\vec{\Theta}=\dprod\limits_{\alpha
=1}^{N}\dprod\limits_{a=1}^{3}\frac{\sqrt{2}}{\left( \sigma _{a}^{\left(
\alpha \right) }\right) ^{2}}d\mu _{a}^{\left( \alpha \right) }d\sigma
_{a}^{\left( \alpha \right) }\text{.}
\end{equation}%
The volume of an extended region $\Delta V_{\mathcal{M}_{s}}\left( \tau 
\text{; }\lambda \right) $ of $\mathcal{M}_{s}$ is defined by,%
\begin{equation}
\Delta V_{\mathcal{M}_{s}}\left( \tau \text{; }\lambda \right) \overset{%
\text{def}}{=}\dprod\limits_{\alpha
=1}^{N}\dprod\limits_{a=1}^{3}\dint\limits_{\mu _{a}^{\left( \alpha \right)
}\left( 0\right) }^{\mu _{a}^{\left( \alpha \right) }\left( \tau \right)
}\dint\limits_{\sigma _{a}^{\left( \alpha \right) }\left( 0\right) }^{\sigma
_{a}^{\left( \alpha \right) }\left( \tau \right) }\frac{\sqrt{2}}{\left(
\sigma _{a}^{\left( \alpha \right) }\right) ^{2}}d\mu _{a}^{\left( \alpha
\right) }d\sigma _{a}^{\left( \alpha \right) }
\end{equation}%
where $\mu _{a}^{\left( \alpha \right) }\left( \tau \right) $ and $\sigma
_{a}^{\left( \alpha \right) }\left( \tau \right) $ are given in $\left(
16\right) $. The quantity that encodes relevant information about the
stability of neighboring volume elements is the the average volume $\bar{V}_{%
\mathcal{M}_{s}}\left( \tau \text{; }\lambda \right) $, 
\begin{equation}
\bar{V}_{\mathcal{M}_{s}}\left( \tau \text{; }\lambda \right) \equiv
\left\langle \Delta V_{\mathcal{M}_{s}}\left( \tau \text{; }\lambda \right)
\right\rangle _{\tau }\overset{\text{def}}{=}\frac{1}{\tau }%
\dint\limits_{0}^{\tau }\Delta V_{\mathcal{M}_{s}}\left( \tau ^{\prime }%
\text{; }\lambda \right) d\tau ^{\prime }\overset{\tau \rightarrow \infty }{%
\approx }e^{3N\lambda \tau }\text{.}
\end{equation}%
This asymptotic regime of diffusive evolution in $\left( 19\right) $
describes the exponential increase of average volume elements on $\mathcal{M}%
_{s}$. The exponential instability characteristic of chaos forces the system
to rapidly explore large areas (volumes) of the statistical manifolds. It is
interesting to note that this asymptotic behavior appears also in the
conventional description of quantum chaos where the entropy increases
linearly at a rate determined by the Lyapunov exponents. The linear entropy
increase as a quantum chaos criterion was introduced by Zurek and Paz $\left[
10\right] $. In our information-geometric approach a relevant variable that
can be useful to study the degree of instability characterizing the ED model
is the information-geometric entropy quantity defined as,%
\begin{equation}
S_{\mathcal{M}_{s}}\overset{\text{def}}{=}\underset{\tau \rightarrow \infty }%
{\lim }\log \bar{V}_{\mathcal{M}_{s}}\left( \tau \text{; }\lambda \right) 
\text{.}
\end{equation}%
Substituting $\left( 18\right) $ in $\left( 19\right) $, equation $\left(
20\right) $ becomes,%
\begin{equation}
S_{\mathcal{M}_{s}}\overset{\tau \rightarrow \infty }{\approx }3N\lambda
\tau \text{.}
\end{equation}%
The entropy-like quantity $S_{\mathcal{M}_{s}}$ in $\left( 21\right) $ is
the asymptotic limit of the natural logarithm of the statistical weight $%
\left\langle \Delta V_{\mathcal{M}_{s}}\right\rangle _{\tau }$ defined on $%
\mathcal{M}_{s}$ and it grows linearly in time, a \textit{quantum} feature
of chaos. \ Indeed, equation $\left( 21\right) $ may be considered the
information-geometric analog of the Zurek-Paz chaos criterion. Zurek and Paz
considered a chaotic system, a single unstable harmonic oscillator
characterized by a potential $V\left( x\right) =-\frac{\lambda x^{2}}{2}$ ($%
\lambda $ is the Lyapunov exponent), coupled to an external environment. In
the \textit{reversible classical limit}, the von Neumann entropy of such a
system increases linearly at a rate determined by the Lyapunov exponent,%
\begin{equation}
S_{\text{quantum}}^{\left( \text{chaotic }\right) }\left( \text{Zurek-Paz}%
\right) \overset{\tau \rightarrow \infty }{\sim }\lambda \tau \text{.}
\end{equation}%
In general, the von Neumann entropy $S_{\text{quantum}}^{\left( \text{chaotic%
}\right) }$ is given by%
\begin{equation}
S_{\text{quantum}}^{\left( \text{chaotic}\right) }=-tr\left( \widehat{\rho }%
\log _{2}\widehat{\rho }\right) =-\sum_{j}\lambda _{j}\log _{2}\lambda _{j}
\end{equation}%
where the normalized $\left( tr\left( \widehat{\rho }\right) =1\right) $
density operator $\widehat{\rho }$ is defined as%
\begin{equation}
\widehat{\rho }=\sum_{j}\lambda _{j}\left\vert \psi _{j}\right\rangle
\left\langle \psi _{j}\right\vert ,\widehat{\rho }\left\vert \psi
_{j}\right\rangle =\lambda _{j}\left\vert \psi _{j}\right\rangle \text{.}
\end{equation}%
Notice that the consideration of $3N$ uncoupled identical unstable harmonic
oscillators characterized by potentials $V_{i}\left( x\right) =-\frac{%
\lambda _{i}x^{2}}{2}$ $\left( \lambda _{i}=\lambda _{j}\text{; }i\text{, }%
j=1\text{, }2\text{,..., }3N\right) $ would simply lead to%
\begin{equation}
S_{\text{quantum}}^{\left( \text{chaotic }\right) }\left( \text{Zurek-Paz}%
\right) \overset{\tau \rightarrow \infty }{\sim }3N\lambda \tau \text{.}
\end{equation}%
The resemblance of equations $\left( 21\right) $ and $\left( 25\right) $,
either in the form or the content is astonishing. A detailed discussion
about this result and an additional discussion about a possible connection
of $\left( 21\right) $ to the Kolmogorov-Sinai entropy, one of the most
powerful indicators of chaos in classical dynamical systems, will appear in $%
\left[ 8\right] $.

\section{EXPONENTIAL DIVERGENCE OF THE JACOBI FIELD INTENSITY ON $\mathcal{M}%
_{s}$}

Finally, we consider the behavior of the one-parameter family of neighboring
geodesics $\mathcal{F}_{G_{\mathcal{M}_{s}}}\left( \lambda \right) \equiv
\left\{ \Theta _{\mathcal{M}_{s}}^{\mu }\left( \tau \text{; }\lambda \right)
\right\} _{\lambda \in 
\mathbb{R}
^{+}}^{\mu =1\text{,..,}6N}$ where,%
\begin{eqnarray}
\mu _{a}^{\left( \alpha \right) }\left( \tau \text{; }\lambda \right) &=&%
\frac{\Lambda ^{2}}{2\lambda }\frac{1}{\cosh \left( 2\lambda \tau \right)
-\sinh \left( 2\lambda \tau \right) +\frac{\Lambda ^{2}}{8\lambda ^{2}}}%
\text{,}  \notag \\
&& \\
\text{ }\sigma _{a}^{\left( \alpha \right) }\left( \tau \text{; }\lambda
\right) &=&\Lambda \frac{\cosh \left( \lambda \tau \right) -\sinh \left(
\lambda \tau \right) }{\cosh \left( 2\lambda \tau \right) -\sinh \left(
2\lambda \tau \right) +\frac{\Lambda ^{2}}{8\lambda ^{2}}}\text{.}  \notag
\end{eqnarray}%
with $\alpha =1$, $2$,$....$, $N$ and $a=1,2,3$. The relative geodesic
spread on a (non-maximally symmetric) curved manifold as $\mathcal{M}_{s}$
is characterized by the Jacobi-Levi-Civita equation, the natural tool to
tackle dynamical chaos $\left[ 11\text{, }12\right] $,%
\begin{equation}
\frac{D^{2}\delta \Theta ^{\mu }}{D\tau ^{2}}+R_{\nu \rho \sigma }^{\mu }%
\frac{\partial \Theta ^{\nu }}{\partial \tau }\delta \Theta ^{\rho }\frac{%
\partial \Theta ^{\sigma }}{\partial \tau }=0
\end{equation}%
where the Jacobi vector field $J^{\mu }$ is defined as,%
\begin{equation}
J^{\mu }\equiv \delta \Theta ^{\mu }\overset{\text{def}}{=}\delta _{\lambda
}\Theta ^{\mu }=\left( \frac{\partial \Theta ^{\mu }\left( \tau \text{; }%
\lambda \right) }{\partial \lambda }\right) _{\tau }\delta \lambda \text{.}
\end{equation}%
Equation $(27)$ forms a system of $6N$ coupled ordinary differential
equations \textit{linear} in the components of the deviation vector field $%
(28)$ but\textit{\ nonlinear} in derivatives of the metric $\left( 7\right) $%
. When the geodesics are neighboring but their relative velocity is
arbitrary, the corresponding geodesic deviation equation is the so-called
generalized Jacobi equation $\left[ 13\right] $. Substituting $\left(
26\right) $ in $\left( 27\right) $ and neglecting the exponentially decaying
terms in $\delta \Theta ^{\mu }$ and its derivatives, integration of $\left(
27\right) $ leads to the following asymptotic expression of the Jacobi
vector field intensity,%
\begin{equation}
J_{\mathcal{M}_{S}}=\left\Vert J\right\Vert =\left( g_{\mu \nu }J^{\mu
}J^{\nu }\right) ^{\frac{1}{2}}\overset{\tau \rightarrow \infty }{\approx }%
3Ne^{\lambda \tau }\text{.}
\end{equation}%
Further details on the derivation of this result for a four-dimensional
statistical manifold are in $\left[ 5\right] $. We conclude that the
geodesic spread on $\mathcal{M}_{s}$ is described by means of an \textit{%
exponentially} \textit{divergent} Jacobi vector field intensity $J_{\mathcal{%
M}_{s}}$, a \textit{classical} feature of chaos. In our approach the
quantity $\lambda _{J}$,%
\begin{equation}
\lambda _{J}\overset{\text{def}}{=}\underset{\tau \rightarrow \infty }{\lim }%
\frac{1}{\tau }\ln \left( \frac{\left\Vert J_{_{\mathcal{M}_{S}}}\left( \tau
\right) \right\Vert }{\left\Vert J_{_{\mathcal{M}_{S}}}\left( 0\right)
\right\Vert }\right)
\end{equation}%
would play the role of the conventional Lyapunov exponents. In conclusion,
we have shown that,%
\begin{equation}
R_{\mathcal{M}_{s}}=-3N\text{, }S_{\mathcal{M}_{s}}\overset{\tau \rightarrow
\infty }{\approx }3N\lambda \tau \text{, }J_{\mathcal{M}_{S}}\overset{\tau
\rightarrow \infty }{\approx }3Ne^{\lambda \tau }\text{.}
\end{equation}%
The Ricci scalar curvature $R_{\mathcal{M}_{s}}$, the information-geometric
entropy $S_{\mathcal{M}_{s}}$ and the Jacobi vector field intensity $J_{%
\mathcal{M}_{S}}$ are proportional to the number of Gaussian-distributed
microstates of the system. This proportionality leads to the conclusion that
there exists a substantial link among these information-geometric measures
of chaoticity, namely%
\begin{equation}
R_{\mathcal{M}_{s}}\sim S_{\mathcal{M}_{s}}\sim J_{\mathcal{M}_{S}}\text{.}
\end{equation}%
Equation $\left( 32\right) $, together with the information-geometric analog
of the Zurek-Paz quantum chaos criterion, equation $\left( 21\right) $,
represent the fundamental results of this work. We believe our theoretical
modelling scheme may be used to describe actual systems where transitions
from quantum to classical chaos scenario occur, but this will be argued
elsewhere $\left[ 8\right] $.

\section{FINAL REMARKS}

In conclusion, a Gaussian ED statistical model has been constructed on a $6N$%
-dimensional statistical manifold $\mathcal{M}_{s}$. The macro-coordinates
on the manifold are represented by the expectation values of microvariables
associated with Gaussian distributions. The geometric structure of $\mathcal{%
M}_{s}$ was studied. The manifold $\mathcal{M}_{s}$ is a curved manifold of
constant negative Ricci curvature $-3N$ . The geodesics of the ED model are
hyperbolic curves on $\mathcal{M}_{s}$. A study of the stability of
geodesics on $\mathcal{M}_{s}$ was presented. The notion of statistical
volume elements was introduced to investigate the asymptotic behavior of a
one-parameter family of neighboring volumes $\mathcal{F}_{V_{\mathcal{M}%
_{s}}}\left( \lambda \right) \equiv \left\{ V_{\mathcal{M}_{s}}\left( \tau 
\text{; }\lambda \right) \right\} _{\lambda \in 
\mathbb{R}
^{+}}$. An information-geometric analog of the Zurek-Paz chaos criterion was
suggested. It was shown that the behavior of geodesics is characterized by
exponential instability that leads to chaotic scenarios on the curved
statistical manifold. These conclusions are supported by a study based on
the geodesic deviation equations and on the asymptotic behavior of the
Jacobi vector field intensity $J_{\mathcal{M}_{s}}$ on $\mathcal{M}_{s}$. A
Lyapunov exponent analog similar to that appearing in the Riemannian
geometric approach to chaos $\left[ 14\right] $ was suggested as an
indicator of chaoticity. We think this is a relevant result since a rigorous
relation among curvature, Lyapunov exponents and Kolmogorov-Sinay entropy is
still under investigation $\left[ 15\right] $ and since there does not exist
a well defined unifying characterization of chaos in classical and quantum
physics due to fundamental differences between the two theories $\left[ 16%
\right] $.

\section{Acknowledgement}

The author is grateful to Dr. Saleem Ali and Adom Giffin for very useful \
comments and suggestions. Special thanks go to Prof. Ariel Caticha for
clarifying explanations on "Entropic Dynamics" and for his constant support
and advice during this work.

\section{References}

\begin{enumerate}
\item A. Caticha, "Entropic Dynamics", AIP Conf. Proc. \textbf{617}, 302
(2002).

\item A. Caticha and A. Giffin, "Updating Probabilities", AIP Conf. Proc. 
\textbf{872}, 31-42 (2006).

\item S. Amari and H. Nagaoka, \textit{Methods of Information Geometry},
Oxford University Press, 2000.

\item C. Cafaro, S. A. Ali and A. Giffin, "An Application of Reversible
Entropic Dynamics on Curved Statistical Manifolds", AIP Conf. Proc. \textbf{%
872}, 243-251 (2006).

\item C. Cafaro and S. A. Ali, "Jacobi Fields on Statistical Manifolds of
Negative Curvature", Physica D (2007), doi: 10.1016/j.physd.2007.07.001.

\item R.A. Fisher, "Theory of statistical estimation" Proc. Cambridge
Philos. Soc. \textbf{122}, 700 (1925).

\item C.R. Rao, "Information and accuracy attainable in the estimation of
statistical parameters", Bull. Calcutta Math. Soc. \textbf{37}, 81 (1945).

\item S. A. Ali and C. Cafaro, "Towards an Information Geometrodynamical
Approach to Classical and Quantum Chaos", accepted for presentation at the "%
\textit{Ettore Majorana Centre}", Erice-Italy (November, 2007).

\item V.I. Arnold, \textit{Mathematical Methods of Classical Physics},
Springer-Verlag, 1989.

\item W. H. Zurek and J. P. Paz, "Quantum Chaos: a decoherent definition",
Physica D\textbf{\ 83}, 300 (1995).

\item M. P. do Carmo, \textit{Riemannian Geometry}, Birkhauser, Boston, 1992.

\item C. W. Misner, K. S. Thorne and J. A. Wheeler, \textit{Gravitation},
Freeman \& Co., San Francisco, 1973.

\item C. Chicone and B. Mashhoon, "The generalized Jacobi equation", Class.
Quantum Grav. \textbf{19}, (2002).

\item L. Casetti et \textit{al}., "Riemannian theory of Hamiltonian chaos
and Lyapunov exponents", Phys. Rev. E \textbf{54}, (1996).

\item T. Kawabe, "Indicator of chaos based on the Riemannian geometric
approach", Phys. Rev. E \textbf{71}, (2005).

\item A. J. Scott et \textit{al}., "Hypersensitivity and chaos signatures in
the quantum baker's map", J. Phys. A \textbf{39}, (2006).
\end{enumerate}

\end{document}